%% file: multiAP.tex
\documentclass[a4paper]{article}
\usepackage{setspace}
\usepackage{fullpage}

\usepackage[utf8]{inputenc}
\usepackage{graphicx}
\usepackage{subcaption}
\usepackage{float}
\usepackage{color, colortbl}
\usepackage{xcolor}
\usepackage{array}
\usepackage{multirow}
\usepackage{footnote}
\usepackage{cite}
\usepackage{lipsum}
\usepackage{threeparttable}
\usepackage{amsmath}
\usepackage{mathtools}
\usepackage[linesnumbered,vlined]{algorithm2e}

\usepackage[hyphens]{url}
\usepackage{hyperref}

\usepackage{soul}

\usepackage{flushend}

\makeatletter
\def\blfootnote{\xdef\@thefnmark{}\@footnotetext}
\makeatother

\usepackage{lipsum}

\begin{document}
\title{Collision-free Operation in High Density WLAN Deployments}
\author{Luis Sanabria-Russo, Boris Bellalta, Nicol\`{o} Facchi, Francesco Gringoli}

\maketitle

\begin{abstract}
WiFi's popularity has led to crowded scenarios composed of many Access Points (AP) and clients, often operating on overlapping channels, producing interference that gravely degrades performance. This misallocation of resources is often the result of multiple WLANs ownership, that is, networks are frequently setup automatically without considering neighbouring APs. In this work we overview the effect of Overlapping BSS (OBSS) from the perspective of the MAC layer, taking special interest on describing the advantages of eliminating collisions with Carrier Sense Multiple Access with Enhanced Collision Avoidance (CSMA/ECA). We propose a single Access Point (AP) and several multi-AP scenarios, including the residential building example proposed for testing the upcoming IEEE 802.11ax amendment. Results using the first NS-3 implementation of CSMA/ECA reveal the advantage of CSMA/ECA's deterministic backoff contention technique, confirming its suitability for very crowded scenarios.
\end{abstract}


\input{introduction}
\input{relatedWork}
\input{eval}
\input{eca}
\input{results}
\input{conclusions}

\section*{Acknowledgements}
This research has been partially supported by the Catalan Government through SGR2014-1173.

\bibliographystyle{plain}
\bibliography{ref}

\end{document}

%% file: introduction.tex
\section{Introduction}\label{intro}
Wireless Local Area Networks (WLANs or WiFi), specified by the IEEE 802.11 standard~\cite{802Standards}, are an almost ubiquitous technology. Its popularity has rendered it as the default wireless access for Local Area Networks (LANs). In fact, we encounter WLANs in public places, coffee shops, offices, stadiums, event halls, and at home. Furthermore, Internet Service Providers (ISP) market WLANs as a business differentiator for bars or restaurants\footnote{Recent Vodafone marketing campaigns offer business services arguing that WLANs increase the flow of clients to the establishment.}.

As the WLAN channel is shared, each user willing to transmit must perform a contention using a carrier sense protocol, called Carrier Sense Multiple Access with Collision Avoidance (CSMA/CA). CSMA/CA is implemented by the Distributed Coordination Function (DCF), which will be referred to as the WLANs's MAC protocol\footnote{CSMA/CA and DCF will be used interchangeably throughout the rest of the text.}. If a node has a packet to transmit, CSMA/CA instructs it to generate a random backoff counter between $[0,W-1]$, where $W$ represents a Contention Window. If the node detects the channel as idle\footnote{Measures the power on the channel to be below a threshold} for an empty slot duration, $\sigma_{e}$, it will decrement its backoff in one. Once the backoff reaches zero the channel should be detected as idle for another period of DIFS before immediately performing a transmission attempt. Values for $\sigma_{e}$ and DIFS are specified for each IEEE 802.11 protocol Physical Layer (PHY)\footnote{$\sigma_{e}=9\mu s; DIFS=34\mu s$, for 802.11 ax~\cite{bellalta2015WCM}}.

Even with the extremely simplified description of CSMA/CA presented above, it is easy to see that WLAN's contention mechanism requires periods of idleness in the channel, as well as the ability to correctly detect this condition. In crowded scenarios, hidden/exposed node problems~\cite{perahia2013next} arise more frequently given that the interference range (carrier sense range) of a node is larger than its communication range~\cite{tuningCarrierSense}. Additionally, Overlapping Basic Service Sets (OBSS) increase the number of transmissions affecting each node's contention (larger contention domain). This condition accentuates the known throughput degradation produced by DCF's random backoff technique, producing throughput starvation in dense WLAN deployments.

CSMA/ECA's deterministic backoff after successful transmissions technique is able to create collision-free schedules, avoid throughput starvation, and achieve higher throughput than CSMA/CA in single WLAN scenarios~\cite{sanabria2014high,ECAqosPaper}. Moreover, the ability to adapt the aforementioned schedule allows CSMA/ECA to increase the number of collision-free users supported, producing considerable reductions in the number of failed transmissions. 

As CSMA/CA performance degrades in OBSS due to a larger contention domain, CSMA/ECA is thought to be a suitable candidate to leverage the effects of collisions and the resulting throughput starvation in crowded OBSS. In this work we overview the effects of crowded scenarios over the performance from the MAC's perspective. We evaluate CSMA/CA and different configurations of CSMA/ECA in order to reveal the consequences of a better collision avoidance mechanism in the these challenging OBSS scenarios. We will review the main subjects involved in the contention for the channel in Section~\ref{background}. We then detail single-AP and multi-AP scenarios in Section~\ref{sim}, while Section~\ref{eca} describes the first NS-3 implementation of CSMA/ECA. Using the NS-3 simulator allows for an accurate configuration of the position of the networking devices, as well as facilitates well-known propagation losses, PHY, and MAC layer software abstractions. Results and conclusions are included in Section~\ref{results} and Section~\ref{conclusions}, respectively.

%% file: relatedWork.tex
\section{Related Work}\label{background}
WLAN's channel is shared, so in order to maximise the available throughput there has to be as much concurrent transmissions as possible. 

As IEEE 802.11 amendments\footnote{IEEE 802.11 n/ac/ax.} propose new functionalities, backwards compatibility mechanisms are enforced to support legacy stations\footnote{Stations using IEEE 802.11g or older specifications a refered to as legacy.}. For instance, all stations in the Basic Service Set (BSS) need to agree which primary channel number they use, bandwidth to operate, as well as being informed about the capabilities supported by the BSS, like frame aggregation or protection mechanisms for ensuring the correct operation of legacy stations. Similarly, Clear Channel Assessment (CCA) carrier sensing threshold values are defined by the standard. These are usually advertised by the Access Point (AP).

\subsection{Carrier sensing thresholds}
CSMA/CA performs two types of carrier sensing for attempting transmissions: Physical Carrier Sensing (PCS), and Virtual Carrier Sensing (VCS). The PCS is done at the Physical Layer (PHY) via the CCA, that is, if after the backoff period the channel energy during a period of DIFS is observed below an Energy Detection (ED), the node will attempt transmission immediately, considering the channel as empty. On the other hand, if the channel is busy the node will keep listening to the channel until it is sensed free for DIFS and then transmits. The other carrier sensing mechanism, VCS, uses control frames to broadcast information about the transmission duration (RTS/CTS mechanism frames~\cite{perahia2013next}). This way other listening stations are aware of ongoing transmissions.

There are several works studying the effects of physical carrier sense sensitivity over the throughput~\cite{DSC-survey}. Further, it is posible to achieve optimal adjustment of the ED threshold so the throughput is maximised in WLANs~\cite{5453868}.

The IEEE 802.11ax High Efficiency WLAN (HEW) Task Group (TGax)~\cite{HEW-scenarios}, in charge of pushing the 802.11 ax amendment has proposed several scenarios and functionalities to be considered for simulation purposes. Furthermore, it focuses on PCS and dynamic ED threshold adaptation to increase spacial reuse~\cite{HEW}, specifically the Dynamic Sensitivity Control (DSC)~\cite{DSC,5453868,UPC-DSC}. Several studies show that by dynamically adjusting the ED threshold it is possible to reduce the effects of neighbouring WLANs that produce throughput degradation in crowded scenarios.

\subsection{CSMA/ECA}
Carrier Sense Multiple Access with Enhanced Collision Avoidance (CSMA/ECA)~\cite{barcelo2008lba,sanabria2014high} is a fully decentralised and collision-free MAC for WLANs. It differs from CSMA/CA in that it uses a deterministic backoff, $B_{\text{d}}=\lceil \text{CW}_{\min}/2\rceil-1$ after successful transmissions, where $\text{CW}_{\min}$ is the minimum contention window of typical value CW$_{\min}=16$. By doing so, contenders that successfully transmitted on a schedule $n$, will transmit without colliding with other successful nodes in future cycles, like in schedule $n+1$. Collision slots being orders of magnitude larger than empty slots degrade the network performance. When CSMA/ECA builds the collision-free schedule all contenders are able to successfully transmit more often, increasing the aggregated throughput beyond CSMA/CA's.

Hysteresis is a property of the protocol that instructs nodes not to reset their backoff stage ($k,k\in[0,\ldots ,m]$) after successful transmissions, but to use a deterministic backoff $B_{\text{d}}=\lceil \text{CW}(k)/2\rceil -1$, where CW$(k)=2^{k}\text{CW}_{\min}$ and CW$_\text{max}=2^{m}\text{CW}_{\min}$. This measure allows the adaptation of the schedule length, admitting many more contenders in a collision-free schedule. To compensate for the different deterministic backoffs that could coexist in a collision-free schedule, Fair Share\footnote{Another property of the protocol.} can be activated. This AMPDU/AMSDU aggregation technique instructs a CSMA/ECA$_\text{Hyst}$ node\footnote{Refering to CSMA/ECA with Hysteresis.} at backoff stage $k$, to transmit $2^{k}$ frames in an AMPDU or AMSDU.

Even if no collision-free schedule is built at an specific moment, CSMA/ECA$_\text{Hyst}$ nodes are always looking for opportunities to reduce the deterministic backoff. The Schedule Reset (SR) mechanism for CSMA/ECA$_\text{Hyst}$ consists in finding the smallest collision-free schedule (if any) between a contender's transmissions and then change the node's deterministic backoff to fit in that schedule~\cite{sanabria2014high}. The \emph{conservative} configuration of the Schedule Reset mechanism consist on filling a bitmap the size of a node's current $B_\text{d}+1$ with the status of each passing slot until the next transmission. Only two states are possible, empty (marked 0) or busy (marked 1). After $\gamma=\frac{\text{CW}_{\max}/2}{B_\text{d} + 1}$ consecutive transmissions, the bitmap is evaluated. If the SR bitmap slots corresponding to any smaller $B_{\text{d}}*$ are found empty, the node will reduce its deterministic backoff, $B_\text{d}\leftarrow B_\text{d}*$. If a collision occurs immediately after the schedule reduction, SR reverses the schedule change before letting CSMA/ECA$_\text{Hyst}$ handle the collision.

Schedule Reset coupled with an increase in the \emph{stickiness}~\cite{barcelo2011tcf} after an effective schedule change has proven to be suitable for noisy scenarios in real hardware implementations of CSMA/ECA~\cite{sanabria2014high}. It is just a simple instruction to the contenders to \emph{stick} to the deterministic backoff even in the event of \emph{stickiness} number of failed transmissions. A default level of stickiness equal to $1$ has proven to provide the better combination of high throughput and low collisions~\cite{ECAqosPaper}. This configuration of CSMA/ECA is referred to as CSMA/ECA$_\text{Hyst+SR}$.

%% file: eval.tex
\section{Scenarios}\label{sim}
\begin{figure}[t]
	\centering
		\includegraphics[width=0.7\linewidth]{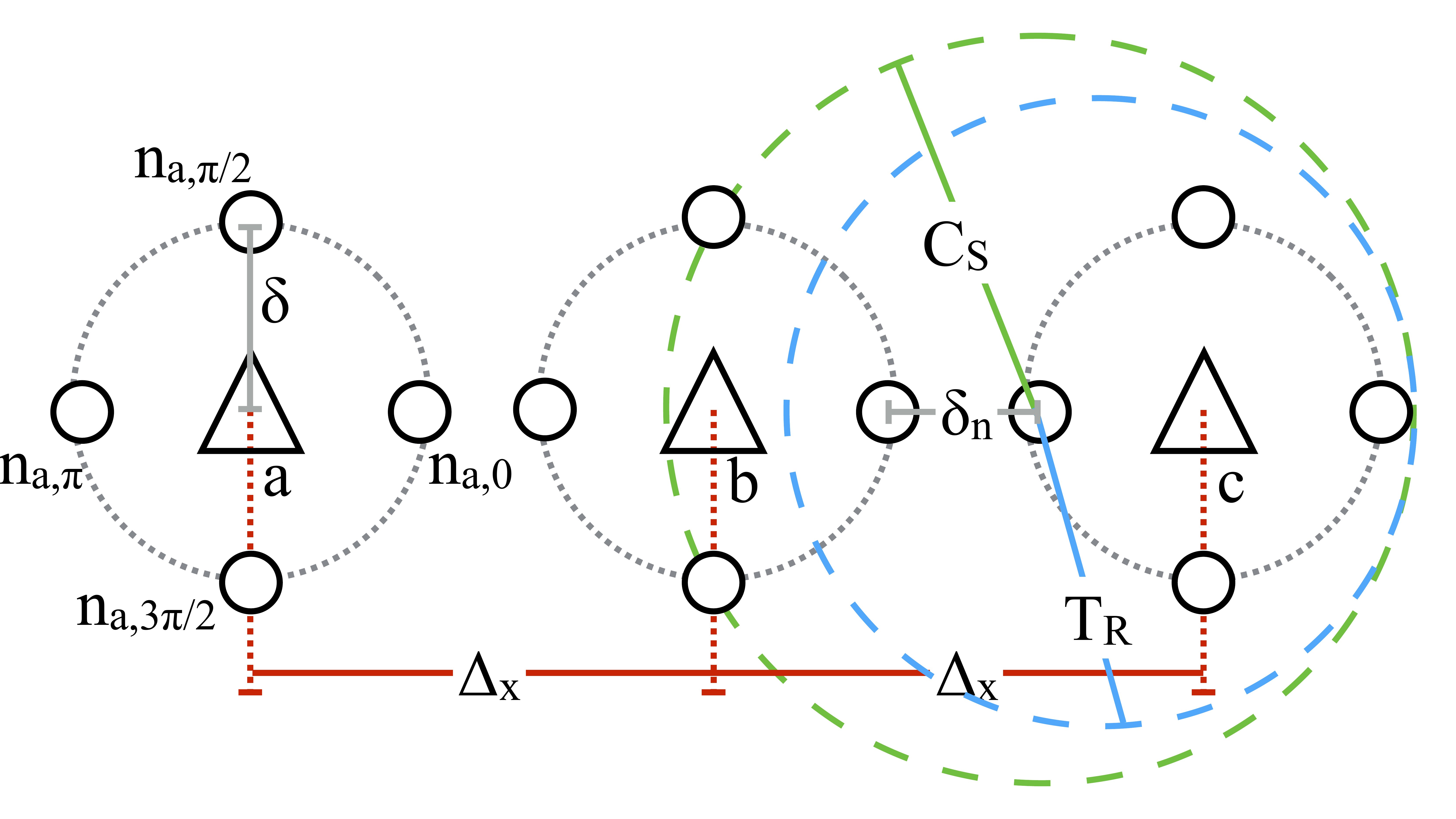}
		\caption{Scenario A example. Solid circles are nodes associated with the AP in the center, represented by a triangle. APs are separated by $\Delta_{x}$ meters, while nodes are at $\delta$ meters from their respective AP. Neighbouring nodes are separated by $\delta_{n}$ meters. Transmissions from nodes within C$_{\text{s}}$, will trigger the carrier sense mechanism, freezing the backoff counter. Nevertheless, only received frames from nodes within T$_{\text{R}}$ are effectively decoded.}
		\label{fig:a}
	\end{figure}

\begin{figure}[t]
	\centering
		\includegraphics[width=0.7\linewidth]{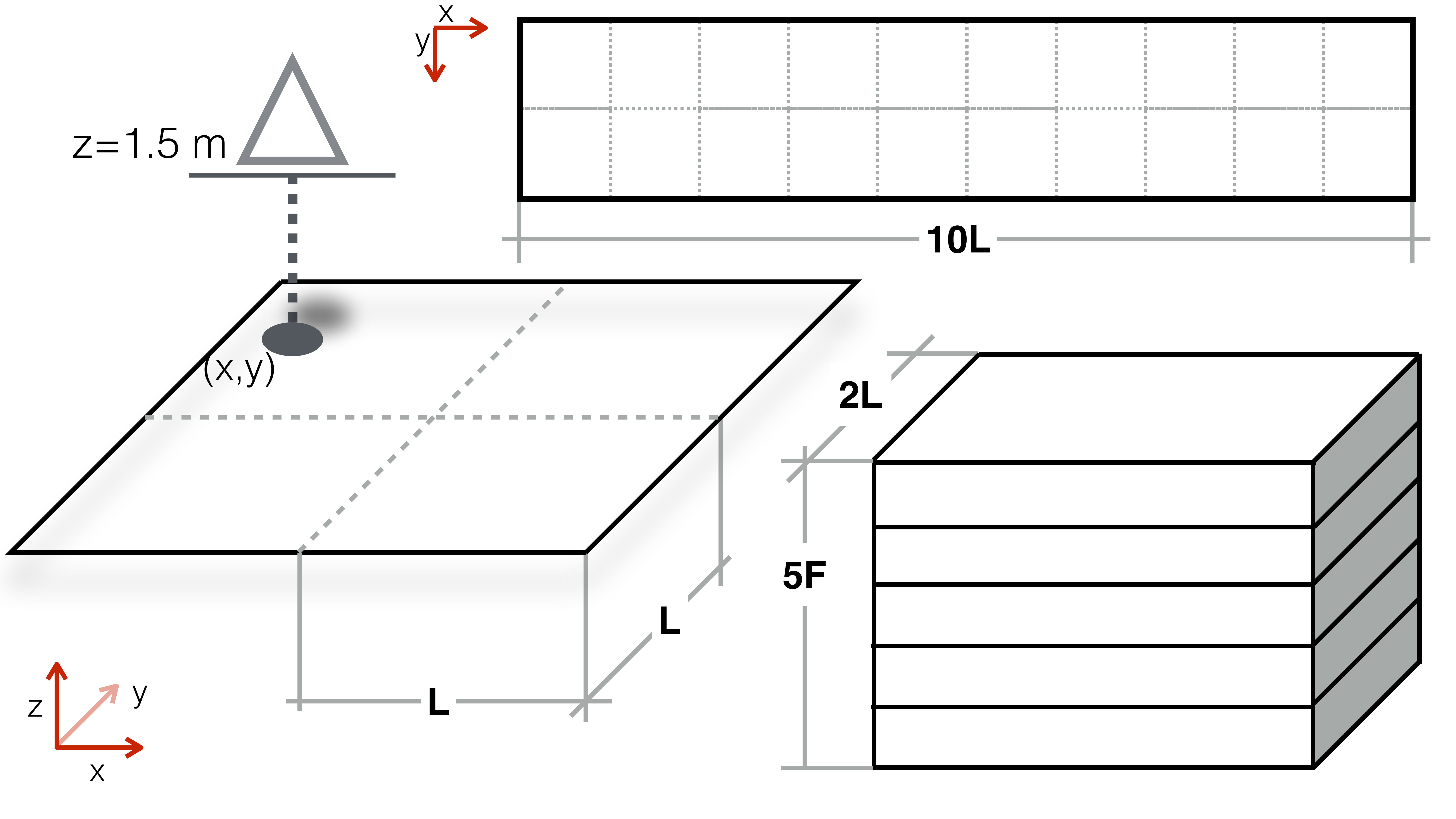}
		\caption{The Scenario HEW represents a common residential building following TGax scenarios~\cite{HEW-scenarios}. The height from floor to ceiling is $F=3$ meters, and $L=10$ meters. APs (represented by the triangle) are randomly placed inside each apartment $q\in[1,\dotsb ,A]$ (of area equals to $L^{2}$) and fixed at $z=1.5$ meters from the floor.}
		\label{building}
	\end{figure}

First, we do a performance evaluation of CSMA/CA, CSMA/ECA and CSMA/ECA$_\text{Hyst+SR}$ in a single AP, fixed rate scenario using NS-3 for the first time (the implementation of CSMA/ECA is described in Section~\ref{eca}). Then, to test the effects of neighbouring WLANs, we define three different scenarios:

	\begin{itemize}
		\item {\bfseries Scenario A:} a linear array of $A$ number of APs, with $N$ nodes forming a circle around each AP $i\in[1,\dotsb,A]$. APs are separated by $\Delta_{x}$ metres, and each node $j\in[0,\dotsb,2\pi]$ associated with AP $i$, that is, node $n_{i,j}$, is at $\delta$ meters from $i$. Neighbouring nodes\footnote{In the Scenario A case with $N=4$ nodes, only $n_{i,0}$ and $n_{k,\pi}$ are considered neighbouring nodes, where $|i-k|=1;~(i,k)\in[1,\dots,A]$.} are separated by $\delta_{n}$. Each node has a Carrier Sense range C$_{\text{s}}$, and a communication range T$_{\text{R}}$. Transmissions from nodes within C$_{\text{s}}$ will trigger the carrier sense mechanism, freezing the backoff counter. Nevertheless, only received frames from nodes within T$_{\text{R}}$ are effectively decoded. Figure~\ref{fig:a} shows and example Scenario A.\\We also test a \emph{control} Scenario A, with no losses suffered inside T$_{\text{R}}=\text{C}_{\text{c}}=2\delta$. Otherwise, Scenario A uses a Log-distance propagation loss model with loss-exponent (details provided below), as proposed by IEEE 802.11ax High Efficiency WLAN (HEW) Task Group (TGax)~\cite{HEW-scenarios}.
		
		\item {\bfseries Scenario B:} AP $i\in[1,\dotsb,A]$ is arranged as in Scenario A, but each node $j\in[0,\dotsb,N]$ is randomly placed at $p_{ij}(x,y)$, where $x,y\in[-\delta,\delta]$ are plane\footnote{It is possible to simulate 3D scenarios by adding an additional coordinate.} coordinates, which is centered in the AP $i$.
		
		\item {\bfseries Scenario HEW:} follows the simulation scenario 1 suggested by TGax~\cite{HEW-scenarios}, also called the residential building scenario. Figure~\ref{building} provides details regarding the dimensions and placement of APs.
	\end{itemize}

Regarding Scenario HEW, the TGax proposes different criteria for evaluating IEEE 802.11ax WLAN. Scenario HEW follows TGax's guidelines in the sense that:
	\begin{itemize}
		\item $L=10$ and $F=3$ are the side of a square room $q\in[1,\dotsb ,A]$ and its height in meters, respectively (see Figure~\ref{building}).
		\item The AP $i\in[1,\dotsb,A]$ is randomly placed inside room $q$ at a fixed heigh of $z=1.5$m.
		\item Nodes also are randomly placed inside room $q$ at a fixed heigh of $z=1.5$m, and associated with AP $i$. We identify a node $j\in[0,\dots ,N-1]$ as $n_{ij}$. $N=10$ per AP.
		\item Walls and floors impose propagation losses to the signal. Our model uses the same propagation loss model and loss-exponents. (\ref{losses}) (based on the one proposed in~\cite{HEW-scenarios}) shows the path loss $p_{l}(x)$ (dB), where $x$ is the known distance to the transmitter, $f_{c}$ is the operating frequency (see Table~\ref{tab1}), $(x>5)$ evaluates to $1$ when the condition is met or returns $0$ otherwise. Finally, $W$ represent the aggregate number of walls traversed to reach the receiver, while $Z$ is the number of floors traversed until the signal arrives at the receiver. All the multi-AP scenarios use this same propagation loss model (only Scenario HEW considers a building, $Z$ and $W$ in~(\ref{losses}) are set to zero otherwise).
	\end{itemize}

	\begin{figure*}[tb]
		\normalsize
			\begin{equation}\label{losses}
					p_{l}(x) = 40.05 + 20~\text{Log}_{10} \left(\frac{f_{c}}{5(10^{9})}\right) + 20~\text{Log}_{10}(\min(x,5)) + (x>5)35~\text{Log}_{10}\left (\frac{x}{5}\right ) + 17Z + 12W
			\end{equation}
	\end{figure*}

We perform tests generating uplink traffic from saturated sources at each node, keeping the MAC queue filled at all times. This means that nodes always have a packet to transmit. Additionally, simulations follow MAC and PHY specifications from the IEEE 802.11n standard, using a $20$ MHz channel in the $5$ GHz band. The rate of the stations is fixed. Details about the CCA and Energy Detection (ED) thresholds, as well as other MAC and PHY details are shown in Table~\ref{tab1}.

	\begin{table}[t]
		\centering
		\caption{PHY, MAC, CCA, and ED parameters used in the simulations}
		\label{tab1}
		\begin{tabular}{|c|c|}
			\hline
			\multicolumn{2}{|c|}{{\bfseries PHY}}\\
			\hline
			{\bfseries Parameter} & {\bfseries Value}\\
			\hline
			PHY rate & $72.2$ Mbps\\
			MCS & HtMcs7\\
			Channel Width & 20~MHz\\
			Operating Frequency $f_{c}$ & 5.24~GHz\\
			Channel Number & 48\\
			Empty slot & $9~\mu s$\\
			DIFS & $34~\mu s$\\
			SIFS & $16~\mu s$\\
			\hline
			\multicolumn{2}{|c|}{{\bfseries MAC}}\\
			\hline
			$\text{CW}_{\min}$ & 16\\
			$\text{CW}_{\max}$ & 1024\\
			Maximum retransmission attempts & 7\\
			Default Packet size (Bytes) & 1470\\
			\hline
			\multicolumn{2}{|c|}{{\bfseries Channel and Tx/Rx properties}}\\
			\hline
			cca1Threshold (CCA) & $-62$ dBm\\
			edThreshold (ED) & $-82$ dBm\\
			Tx power & $15$ dBm\\
			Per wall losses & $12$ dB\\
			Per floor losses & $17$ dB\\
			\hline
		\end{tabular}
	\end{table}

%% file: eca.tex
\section{CSMA/ECA NS-3 Implementation}\label{eca}
The backoff mechanism controlled by the \texttt{EdcaTxopN} class is modified to react differently to the effective reception of an ACK. That is, instead of following CSMA/CA backoff mechanism, nodes are reconfigured to follow CSMA/ECA$_\text{Hyst+SR}$ upon the call to the \texttt{EdcaTxopN::GotAck} method\footnote{Which happens everytime a successful transmission is acknoledged by the receiver.}.

The Schedule Reset mechanism involved the modification of the \texttt{DcfManager} class. It creates a bitmap according to the current CSMA/ECA$_\text{Hyst+SR}$ deterministic backoff, and updates it following the channel conditions at each decrementing slot.

\subsection*{Position of the nodes}
The scenarios defined in Section~\ref{sim} are implemented using different mobility and propagation loss models provided by NS-3. Specifically: 

\begin{itemize}
  \item Each node's position is fixed during simulation time (use \texttt{ConstantPositionMobilityModel} class).
  \item The signal is attenuated according to~(\ref{losses}) with the help of the\\ \texttt{ThreeLogDistancePropagationLossModel}~\cite{stoffers2012comparing,DSCNS3}. 
\end{itemize} 

The design and configuration of a building is made simple by the \texttt{Building} class in NS-3~\cite{ns3BuildingDesing}. It provides several sub-classes and methods for specifying size, materials and attenuation properties using the \texttt{HybridBuildingsPropagationLossModel} class on top of the \texttt{ThreeLogDistancePropagationLossModel}.

Our implementation was made using the NS-3~\cite{ns3} network simulator, and can be accessed via~\cite{CSMA-ECA-NS3}. A tutorial on how to use CSMA/ECA MAC for WiFi in NS-3 is provided by~\cite{eca-ns3-tutorial}.

%% file: results.tex
\section{Results}\label{results}
We proceed to do a series of performance evaluations using the aforementioned scenarios, modifying its characteristics in order to understand the behaviour of CSMA/CA (DCF), CSMA/ECA, and different configurations of CSMA/ECA with Hysteresis and the Schedule Reset Mechanism (CSMA/ECA$_\text{Hyst+SR}$). If not specified otherwise, results are derived from five iterations of a twenty five second NS-3 simulation with different seeds.

\subsection{Single AP}
	\begin{figure*}[tb]
	\centering
		\includegraphics[width=0.3\linewidth, angle=-90]{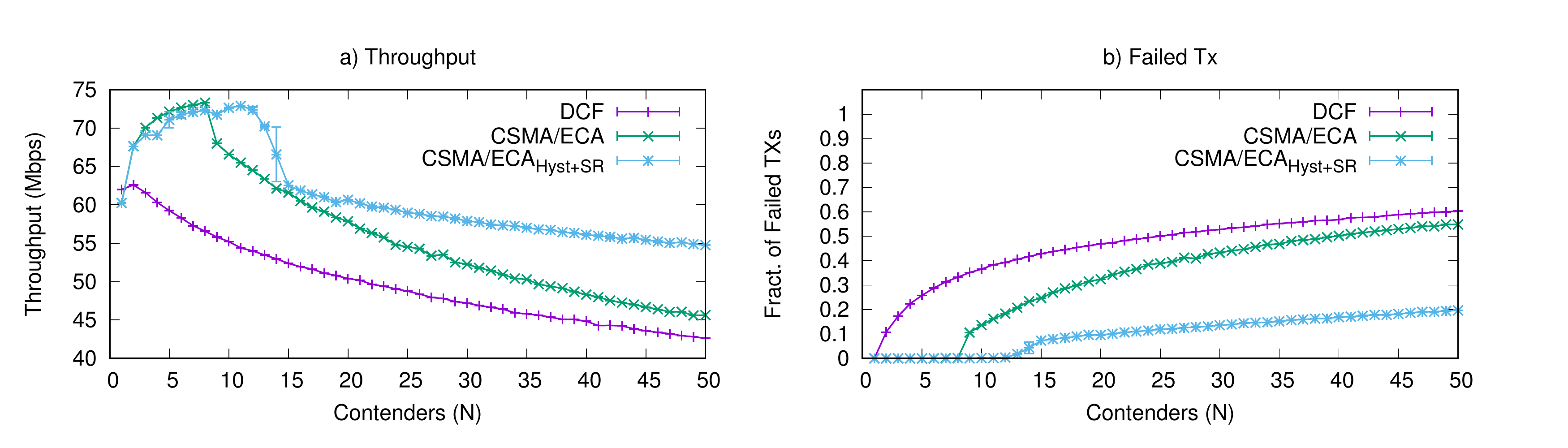}
		\caption{a) Average aggregate throughput and b) Average fraction of failed transmissions for DCF, CSMA/ECA and CSMA/ECA$_\text{Hyst+SR}$ in saturation. Single AP}
		\label{single:throughput}
	\end{figure*}

Figure~\ref{single:throughput} shows: a) average aggregate throughput and b) average fraction of failed transmissions under saturated traffic conditions. The scenario of this test supposes perfect communication among all nodes. Results for DCF, CSMA/ECA, and CSMA/ECA with Hysteresis and conservative Schedule Reset with dynamic stickiness, namely CSMA/ECA$_\text{Hyst+SR}$, are presented.

Results show how DCF's throughput degrades as the number of contenders increases. This is due to the channel time wasted recovering from collisions. On the other hand, when $N\leq B_{\text{d}}$, CSMA/ECA is able to reach collision-free operation. Further, applying Hysteresis allows CSMA/ECA$_\text{Hyst+SR}$ to increase the size of the collision-free schedule augmenting the overall throughput for a greater number of contenders. Schedule Resets seeks opportunities to reduce the size of the deterministic backoff to prevent large periods between successful transmissions. 

\subsection{Multi-AP Scenarios}
	\begin{figure*}[tb]
	\centering
		\includegraphics[width=0.28\linewidth, angle=-90]{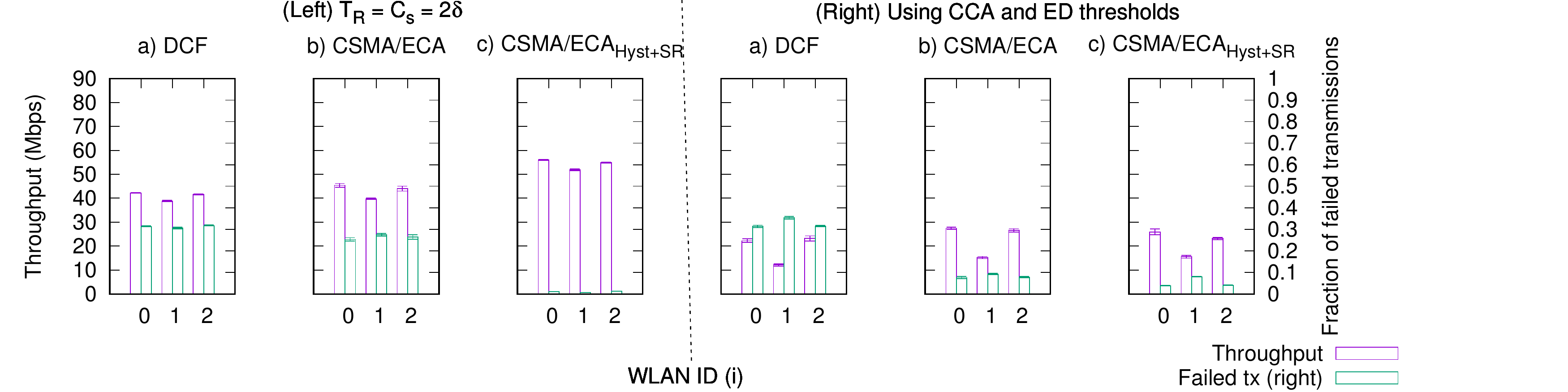}
		\caption{Overall aggregate throughput and fraction of failed transmissions for a) DCF, b) CSMA/ECA, and c) CSMA/ECA$_\text{Hyst+SR}$. Left side of the figure shows the control Scenario A, where $T_{\text{R}}=C_{\text{s}}=2\delta$. In the right side of the figure $T_{\text{R}}$ and $C_{\text{s}}$ follow CCA and ED thresholds}
		\label{fig1}
	\end{figure*}
	
A control Scenario A is shown on the left of Figure~\ref{fig1}, where the legend is located at the bottom right corner of the figure. Here, $T_{\text{R}}=C_{\text{s}}=2\delta$ m, so the effect of neighbouring nodes' transmissions can be delimited with precision. This Scenario A configuration implies that the transmissions from a border node $n_{a,0}$ will trigger neighbouring node $n_{b,\pi}$ and AP $b$'s CCA mechanism, deferring their transmissions. 

On the other side, at the right of Figure~\ref{fig1} $T_{\text{R}}$ and $C_{\text{s}}$ ranges depend on the received signal power, that is, are subject to the CCA and ED thresholds, thus affected by propagation losses. Results in the figure are derived with $N=4$, $\Delta_{x}=15$ m, $\delta=\frac{1}{3}\Delta_{x}$ m, and $\delta_{n}=\delta$. 
	
\begin{figure}[tb]
	\centering
		\includegraphics[width=0.4\linewidth, angle=-90]{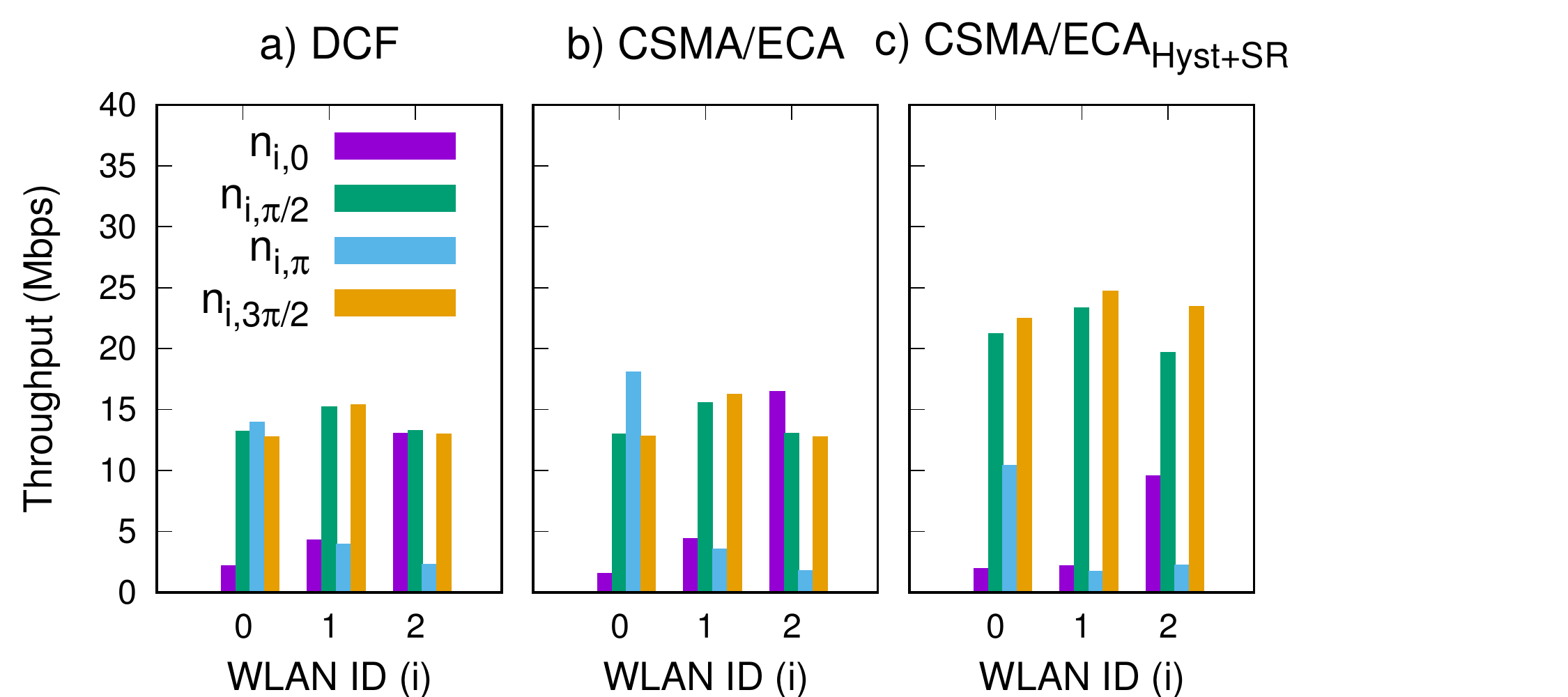}
		\caption{Throughput per station for a) DCF, b) CSMA/ECA, and c) CSMA/ECA$_\text{Hyst+SR}$ in the control Scenario A shown in the left side fo Figure~\ref{fig1}}
		\label{fig2}
	\end{figure}

Figure~\ref{fig1} (left) shows how the throughput is degraded in the middle WLAN-1. This is caused by the transmissions of nodes from adjacent networks. For instance, when a neighbouring node from WLAN-0, $n_{0,0}$ transmits, $n_{1,\pi}$ and AP-1 detect the channel as busy. Now, suppose that $n_{0,0}$ and $n_{2,\pi}$ transmit at the same time. This condition will completely prevent WLAN-1 nodes from transmitting successfully during the aforementioned nodes' transmissions, bringing periods of inactivity that contribute to the observed throughput degradation. Figure~\ref{fig2} clearly shows that neighbouring nodes are the most negatively affected, supporting our assumptions over the control Scenario A.

Still focusing on Figure~\ref{fig1} (left), CSMA/ECA shows higher fraction of failures than CSMA/ECA$_\text{Hys+SR}$. This is because Hysteresis allows larger schedules, avoiding collisions more efficiently. DCF nodes on the other hand waste channel time recovering from collisions.

Using the default CCA and ED thresholds increases $T_{\text{R}}$ and $C_{\text{s}}$ when compared to the control Scenario A. Therefore, for an example node $n_{ij}$ there will be more transmissions triggering the CCA mechanism, deferring transmissions. As mentioned in~\cite{jamil2014improving}, there are now more nodes in the \emph{contention domain}, increasing the collision probability. Looking at the right side of Figure~\ref{fig1}, it shows a higher number of failed transmissions in DCF, coupled with a considerable overall throughput degradation. CSMA/ECA$_\text{Hyst+SR}$ on the other hand, shows less failures, mainly due to a better collision avoidance mechanism.

As CSMA/ECA$_\text{Hyst+SR}$ nodes rapidly reach large deterministic backoffs, they are able to produce collision-free schedules with enough empty slots, leveraging the effects of neighbouring nodes' transmissions. CSMA/ECA on the other hand still shows higher fraction of losses, but the periods of scheduled collision-free operation prevent further increase. Finally, DCF is gravely affected by the higher collision probability. 

\begin{figure}[tb]
	\centering
		\includegraphics[width=0.4\linewidth, angle=-90]{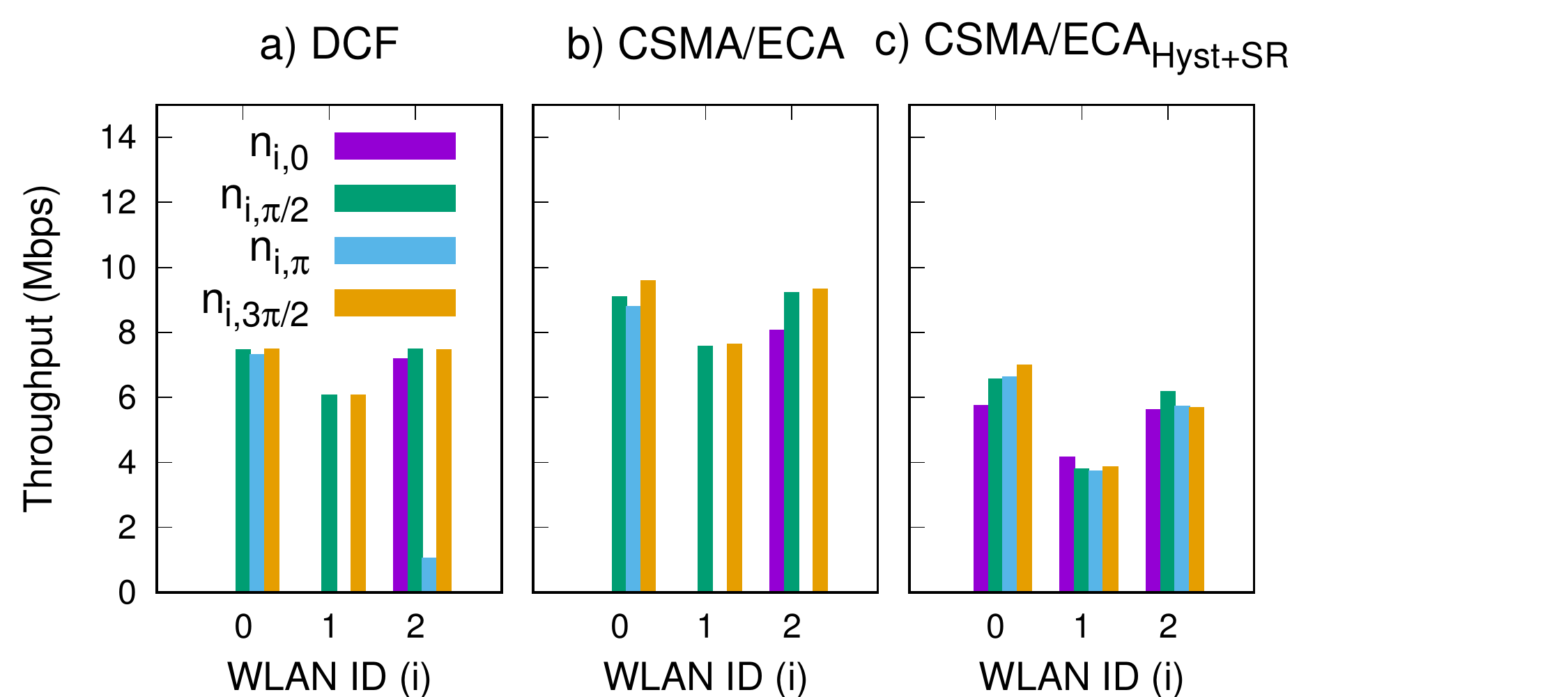}
		\caption{Throughput per station for a) DCF, b) CSMA/ECA, and c) CSMA/ECA$_\text{Hyst+SR}$, following the right side of Figure~\ref{fig1}. That is, using Scenario A layout of the nodes, but following CCA and ED thresholds to determine $T_{\text{R}}$ and $C_{\text{s}}$ ranges}
		\label{fig3}
	\end{figure}

Figure~\ref{fig3} shows the throughput per station. Results from both Figure~\ref{fig1} (right) and Figure~\ref{fig3} suggest this is a very collision-prone scenario, and CSMA/ECA$_\text{Hyst+SR}$ is able to avoid starvation, distributing the available throughput of each WLAN more efficiently than DCF, or CSMA/ECA.

\subsection{Scenario B and many more users}\label{scenarioB}
	\begin{figure*}[tb]
	\centering
		\includegraphics[width=0.35\linewidth, angle=-90]{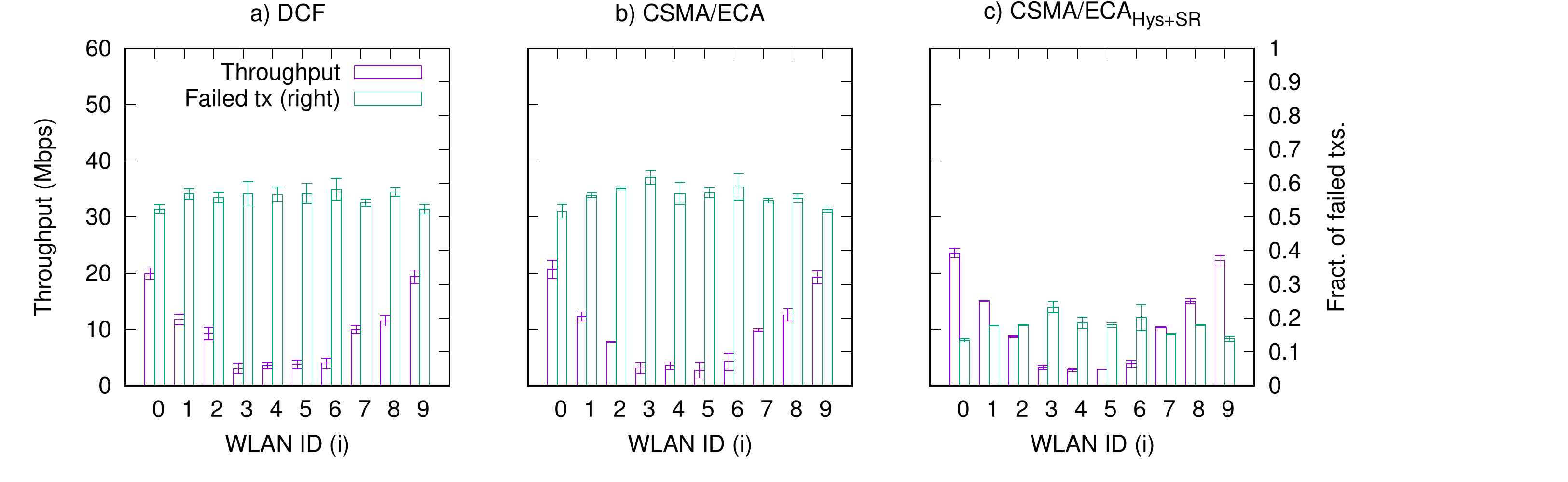}
		\caption{Shows the overall aggregate throughput and fraction of failed transmissions for a) DCF, b) CSMA/ECA, and c) CSMA/ECA$_\text{Hyst+SR}$. It follows Scenario B, where $N=20$ nodes are randomly located around $A=10$ APs, in saturation using Basic Access}
		\label{fig4}
	\end{figure*}

This section presents results for Scenario B with $N=20$ and $A=10$. That is, node $i\in[1,\dotsb,N]$ are randomly placed around AP $i\in[1,\dotsb,A]$ at $p_{ij}(x,y)$, where $x,y \in [-\delta,\delta]$, and elevated $z=1.5$m from the floor. As before, $\Delta_{x}=15$ m, and $\delta=\frac{1}{3}\Delta_{x}$ m.

\begin{figure}[tb]
	\centering
		\includegraphics[width=0.5\linewidth, angle=-90]{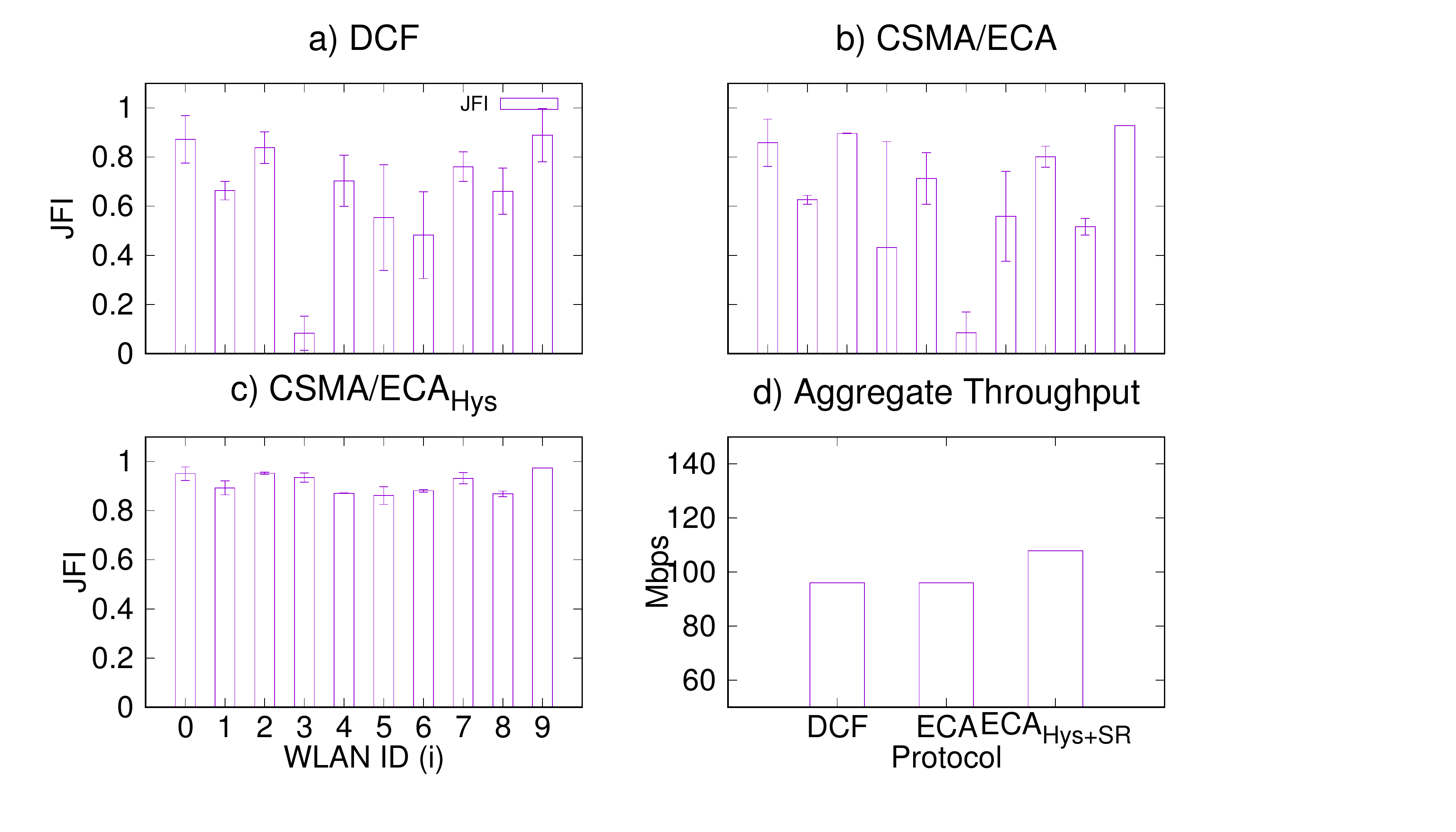}
		\caption{JFI~\cite{JFI} for Scenario B simulations (see Figure~\ref{fig4}). Shows a) DCF, b) CSMA/ECA, c) CSMA/ECA$_\text{Hyst+SR}$, and d) Aggregate throughput for the three tests performed with Scenario B.}
		\label{fairness}
	\end{figure}

Figure~\ref{fig4} shows that CSMA/ECA$_\text{Hyst+SR}$ is more efficient at reducing the fraction of failed transmissions. Additionally, Figures~\ref{fairness}a-c show the JFI~\cite{JFI} for all WLANs using one of the three tested protocols: DCF, CSMA/ECA and CSMA/ECA$_\text{Hyst+SR}$. Results indicate that CSMA/ECA$_\text{Hyst+SR}$ not only is able to increase the fairness among contenders of the same WLAN, but also provides an aggregate throughput increase as a consequence. This can be observed in Figure~\ref{fairness}d.

\subsection{Scenario HEW: residential building}

\begin{figure*}[tb]
	\centering
		\includegraphics[width=0.39\linewidth, angle=-90]{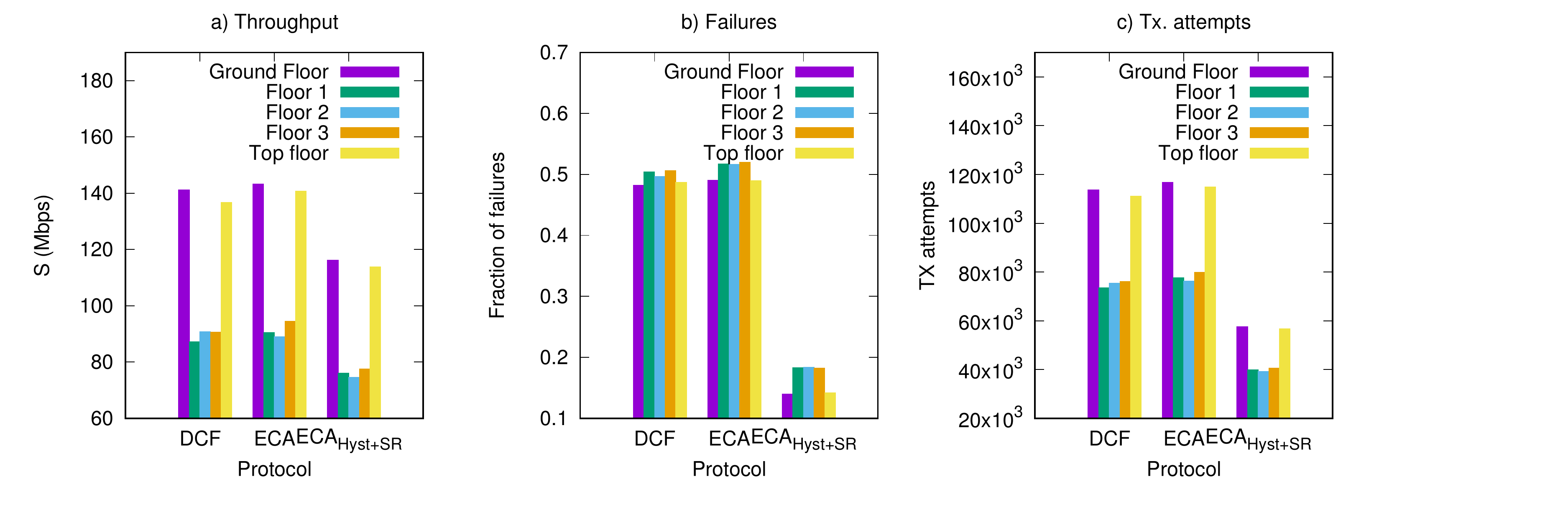}
		\caption{Scenario HEW results, showing a) throughput (S), b) fraction of failed transmissions, and c) number of transmission attempts. Each figure shows the aggregate metric per floow, please refer to Figure~\ref{building} for orientation. The walls and floors impose propagation losses presented in Table~\ref{tab1}}
		\label{figCthroughput}
	\end{figure*}

Figure~\ref{figCthroughput} gathers the aggregate results for the Scenario HEW or residential building simulations, with $N=10$ nodes per WiFi. Figure~\ref{figCthroughput}a shows the aggregate throughput per floor in the example building of Figure~\ref{building}. As expected, the bottom and top floors have a smaller contention domain, showing higher throughput due to a lower fraction of losses in Figure~\ref{figCthroughput}b. CSMA/ECA$_\text{Hyst+SR}$ (ECA$_\text{Hyst+SR}$ in Figure~\ref{figCthroughput}) stations are unable to reduce the deterministic backoff, ending with a big period between successful transmissions which translates in a lower overall throughput. Nevertheless, Figure~\ref{figCthroughput}b shows that it is very effective at reducing failures.

Figure~\ref{fairnessC} shows overall metrics of throughput (S), JFI, failures, and transmission attempts. Results show higher throughput for CSMA/ECA over DCF, despite having around 50\% of transmissions resulting in failure. The use of a deterministic backoff after successful transmissions creates periods of schedule-like transmissions (as in~\cite{BECA-test, CF-MAC}), increasing the number of successful transmissions.

Attempting to increase the aggressiveness of CSMA/ECA$_\text{Hyst+SR}$ may enhance the aggregate throughput for this protocol. In Figure~\ref{ECAcomp} we show different configurations of CSMA/ECA$_\text{Hyst+SR}$, namely: Hysteresis only (\emph{Hyst}), Schedule Reset as in previous tests (\emph{Hyst+SR}), SR with a reduced $\text{CW}_{\max}\leftarrow 255$ (\emph{Hyst+SR$_\text{R}$}), and a configuration of SR denoted as aggressive\footnote{It simply means $\gamma = 1$ and a \emph{halving} of the schedule.} (\emph{Hyst+SR aggr.}). Even-though \emph{Hyst+SR$_\text{R}$} shows higher throughput, it increases the fraction of failed transmissions. The aggressive schedule halving represented by \emph{Hyst+SR aggr.} in Figure~\ref{ECAcomp} is not able to reduce the schedule to lower values, despite just analysing the bitmap after 2 consecutive transmissions ($\gamma = 1$). We select \emph{Hyst+SR} as the reference protocol because is the configuration that provides better tradeoff between overall aggregate throughput and fraction of failures.

\subsubsection*{Softening the conditions using different channels} having a very big contention domain increases the percentage of failed transmissions, nevertheless, CSMA/ECA$_\text{Hyst+SR}$ is able to leverage this issue using a deterministic backoff and Hysteresis. Despite being outperformed by CSMA/ECA in the residential building scenario, results from Scenario B in Section~\ref{scenarioB} show that CSMA/ECA$_\text{Hyst+SR}$ provides a considerable reduction of failures while increasing the overall throughput and fairness. This can be beneficial for applications where low losses and fairness are preferred. Furthermore, as less transmission attempts are performed CSMA/ECA$_\text{Hyst+SR}$ may constitute an advantage for energy constrained applications.

The following presents simulation results using Scenario HEW in saturation with a different WiFi channel for each room of the building. First, a distribution using only $C=8$ non-overlapping WiFi channels in the IEEE 802.11n 5GHz band (shown in Figure~\ref{channelDist}), and then a more efficient distribution using $C=20$ (see Figure~\ref{channelDist-2}).

\begin{enumerate}

\item Figure~\ref{channelDist} shows the two types of WiFi channel assignments for each floor of the building in Figure~\ref{building} using $C=8$ non-overlapping channels. Floors 0, 2 and 4 use TypeA, while floors 1 and 3 use TypeB.

		\begin{figure}[H]
	\centering
		\includegraphics[width=0.3\linewidth, angle=-90]{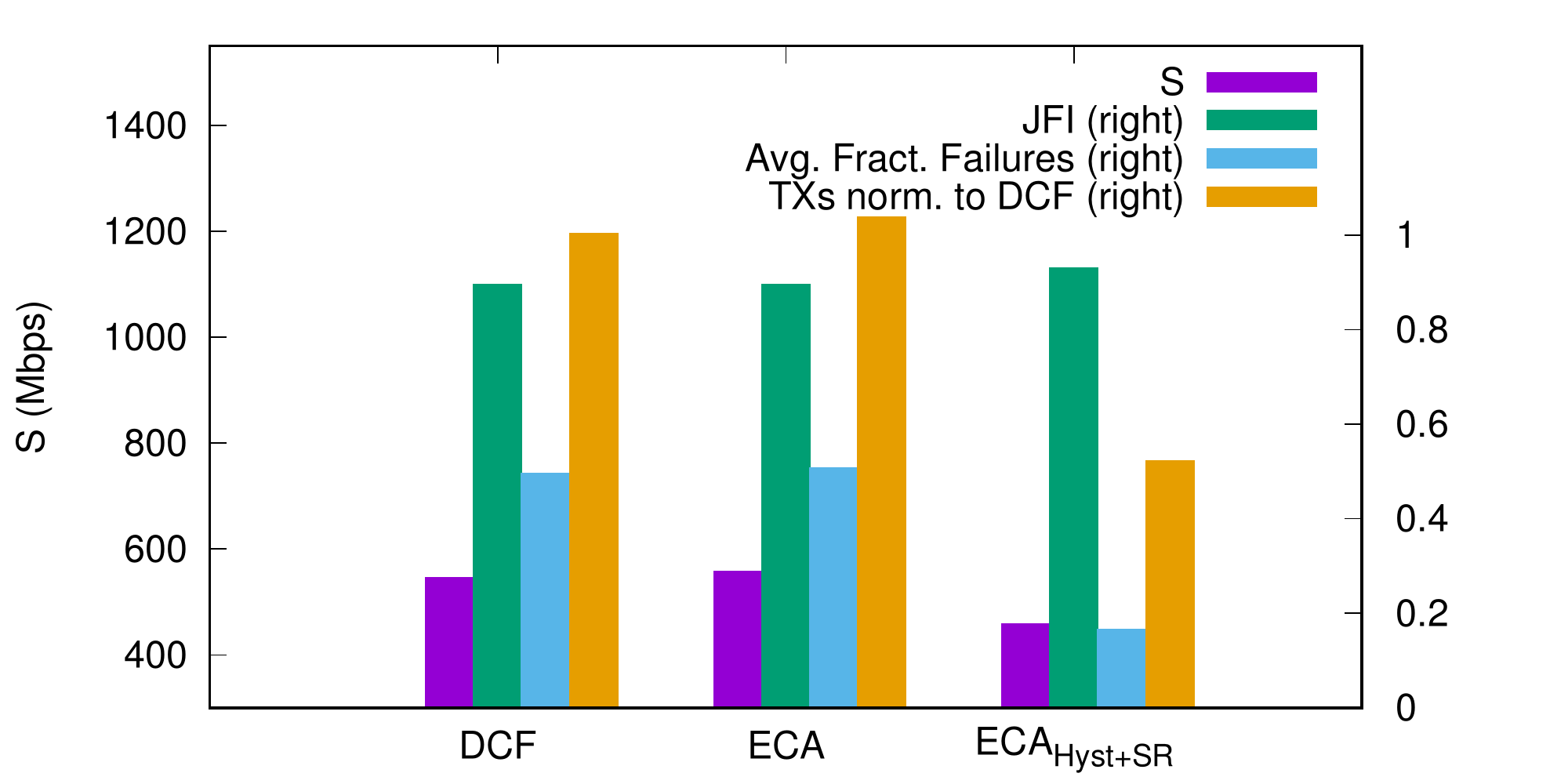}
		\caption{Overall aggregate throughput (S), JFI~\cite{JFI}, and average aggregated fraction of failed transmissions for all protocols in Scenario HEW.}
		\label{fairnessC}
	\end{figure}

	\begin{figure}[H]
	\centering
		\includegraphics[width=0.3\linewidth, angle=-90]{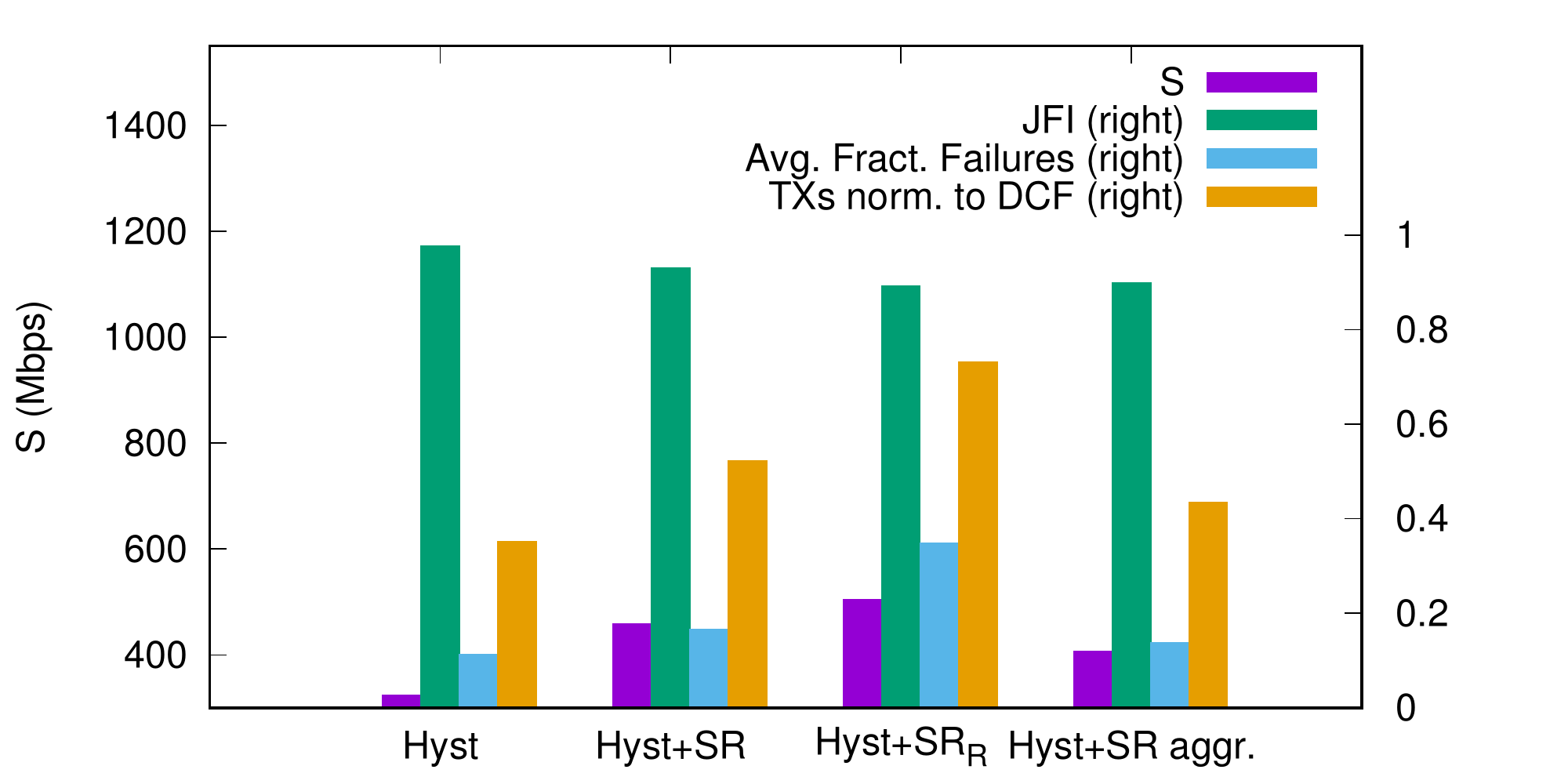}
		\caption{Comparison among different CSMA/ECA$_\text{Hyst+SR}$ configurations.}
		\label{ECAcomp}
	\end{figure}

	\begin{figure}[tb]
	\centering
		\includegraphics[width=0.55\linewidth]{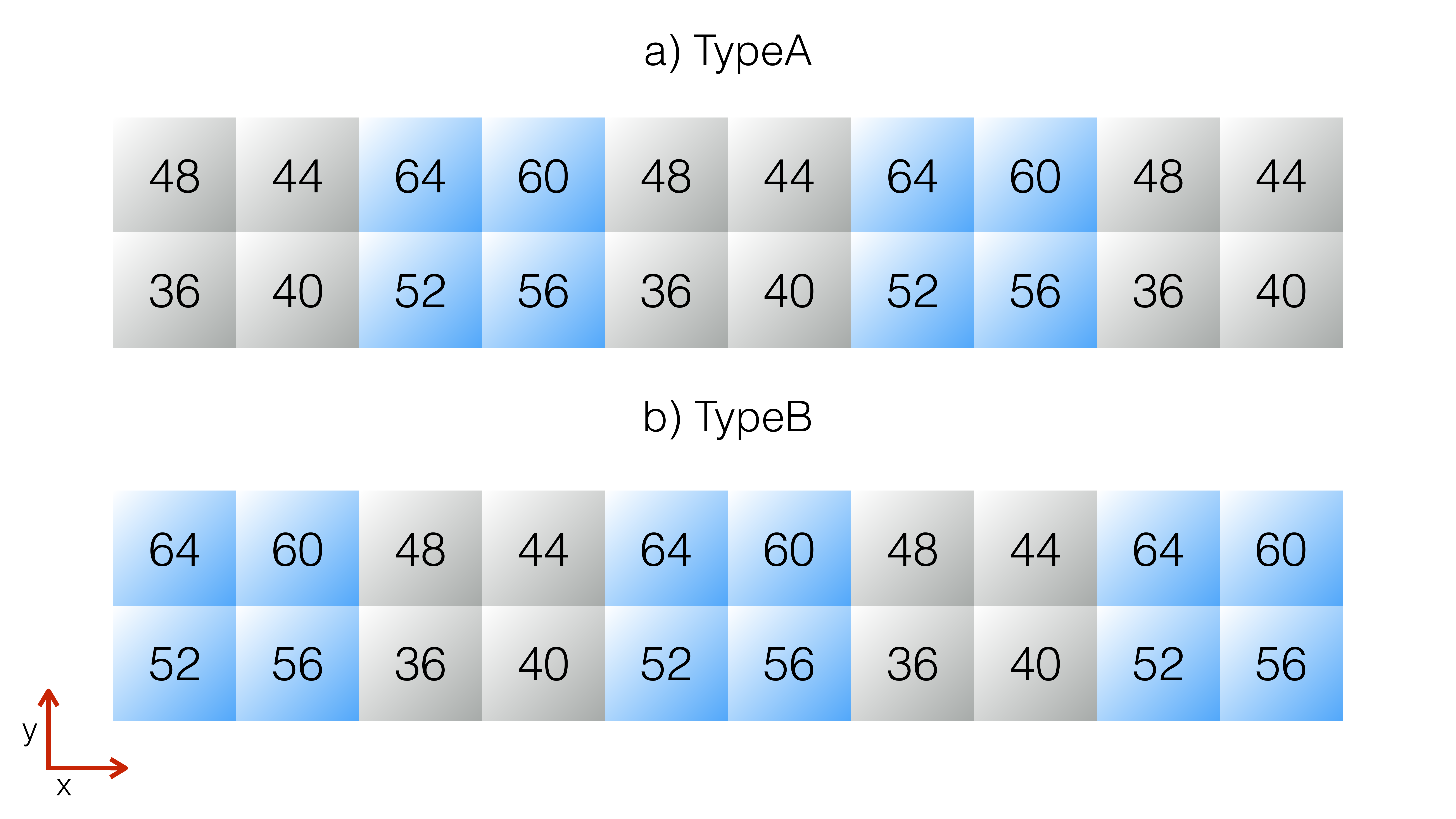}
		\caption{a) TypeA: WiFi channel allocations for floors 0, 2 and 4. b) TypeB: WiFi channel allocations for floor 1 and 3. Working in the IEEE 802.11n 5GHz band and using $C=8$ non-overlapping channels. Please refer to Figure~\ref{building} for orientation}
		\label{channelDist}
	\end{figure}
	
	\begin{figure}[tb]
	\centering
		\includegraphics[width=0.39\linewidth, angle=-90]{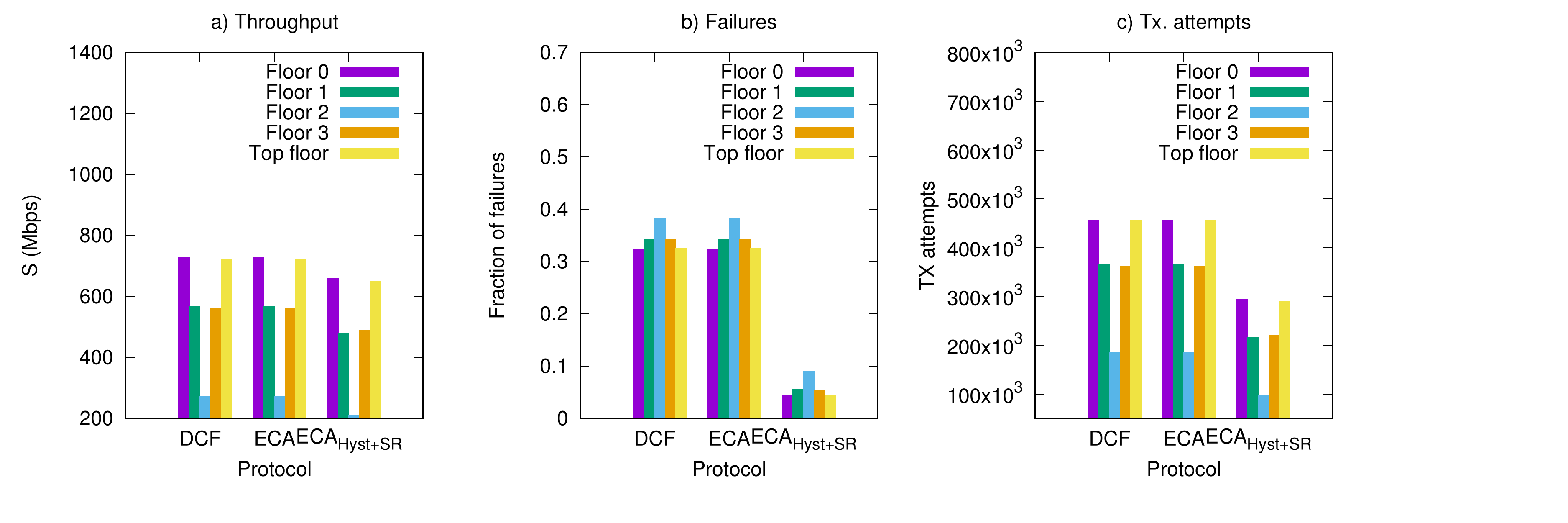}
		\caption{Scenario HEW with efficient channel allocation results using $C=8$ non-overlapping channels, showing a) throughput (S), b) fraction of failed transmissions, and c) number of transmission attempts. Each figure shows the aggregat metric per floow, please refer to Figure~\ref{building} for orientation. The walls and floors impose propagation losses presented in Table~\ref{tab1}}
		\label{figCthroughput-segmented}
	\end{figure}

Figure~\ref{figCthroughput-segmented}, shows the results obtained when allocating different channels for each WLAN. Interestingly, the effect over the middle floor is easily observed, revealing higher failures and lower throughput than the rest. CSMA/ECA and DCF show similar performance, as $N>B_{\text{d}}$ this is expected. On the other hand, CSMA/ECA$_\text{Hyst+SR}$ nodes are unable to reduce the schedule length any further, ending with big periods between successful transmissions that translate into lower throughput. Nevertheless, the efficient collision avoidance mechanisms use by this protocol reduces the fraction of failures and transmissions attempts considerably.\\

\item Then, we proceed to an even more efficient allocation of the available non-overlapping channels using $C=20$. The channel distribution is shown in Figure~\ref{channelDist-2}, while results are presented in Figure~\ref{figCthroughput-segmented-big}.

As each WLAN contention domain is effectively reduced by a more efficient distribution of the non-overlapping channels, a fairness and throughput increase is evidenced. Further, the middle floor (Floor 2) is not specially affected.

In this scenario CSMA/ECA$_\text{Hyst+SR}$ is still unable to outperform CSMA/ECA. Nevertheless, the same benefits in terms of failed transmissions and transmission attempts are observed.

	\begin{figure}[tb]
	\centering
		\includegraphics[width=0.7\linewidth]{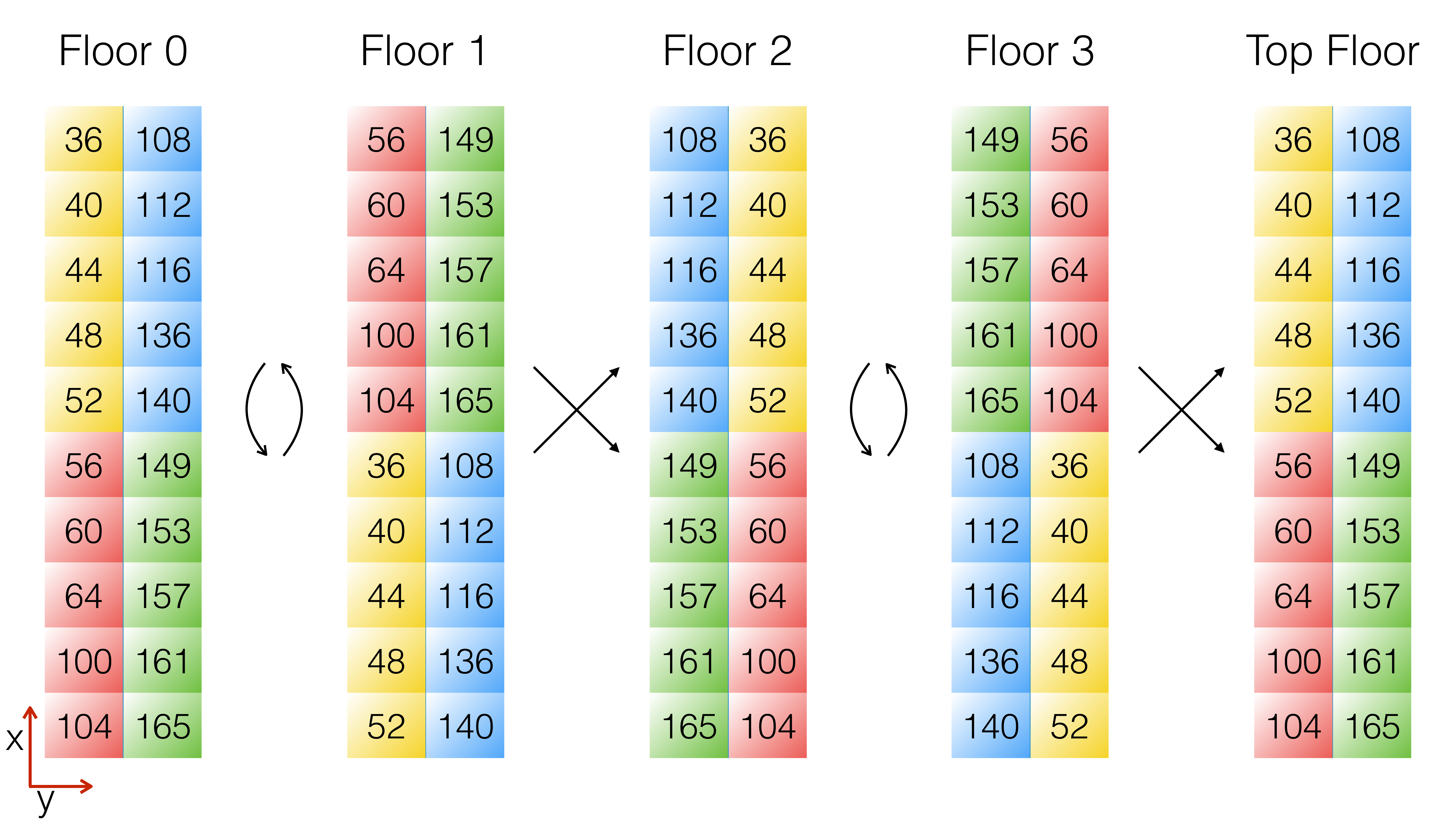}
		\caption{Channel allocation for the building shown in Figure~\ref{building}, using $C=20$ non-overlapping channels in the IEEE 802.11n 5GHz band}
		\label{channelDist-2}
	\end{figure}

	\begin{figure}[tb]
	\centering
		\includegraphics[width=0.39\linewidth, angle=-90]{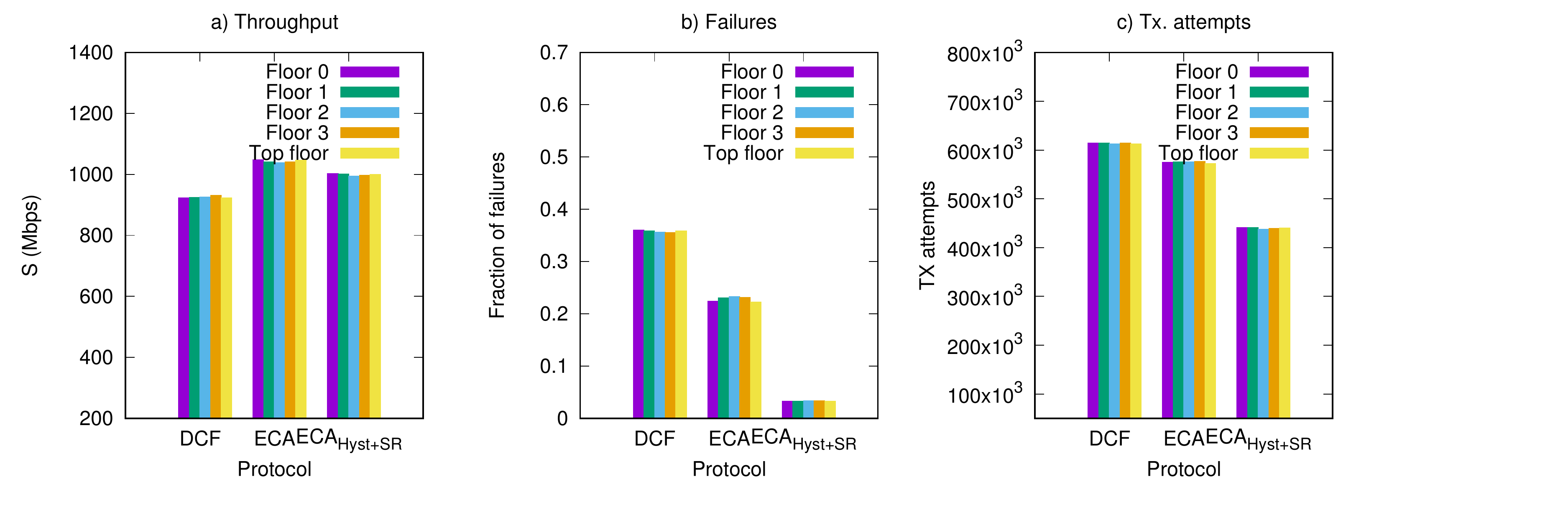}
		\caption{Scenario HEW with efficient channel allocation results using 20 non-overalapping channels, showing a) throughput (S), b) fraction of failed transmissions, and c) number of transmission attempts. Each figure shows the aggregat metric per floow, please refer to Figure~\ref{building} for orientation. The walls and floors impose propagation losses presented in Table~\ref{tab1}}
		\label{figCthroughput-segmented-big}
	\end{figure}
	
		\begin{figure}[tb]
	\centering
		\includegraphics[width=0.39\linewidth, angle=-90]{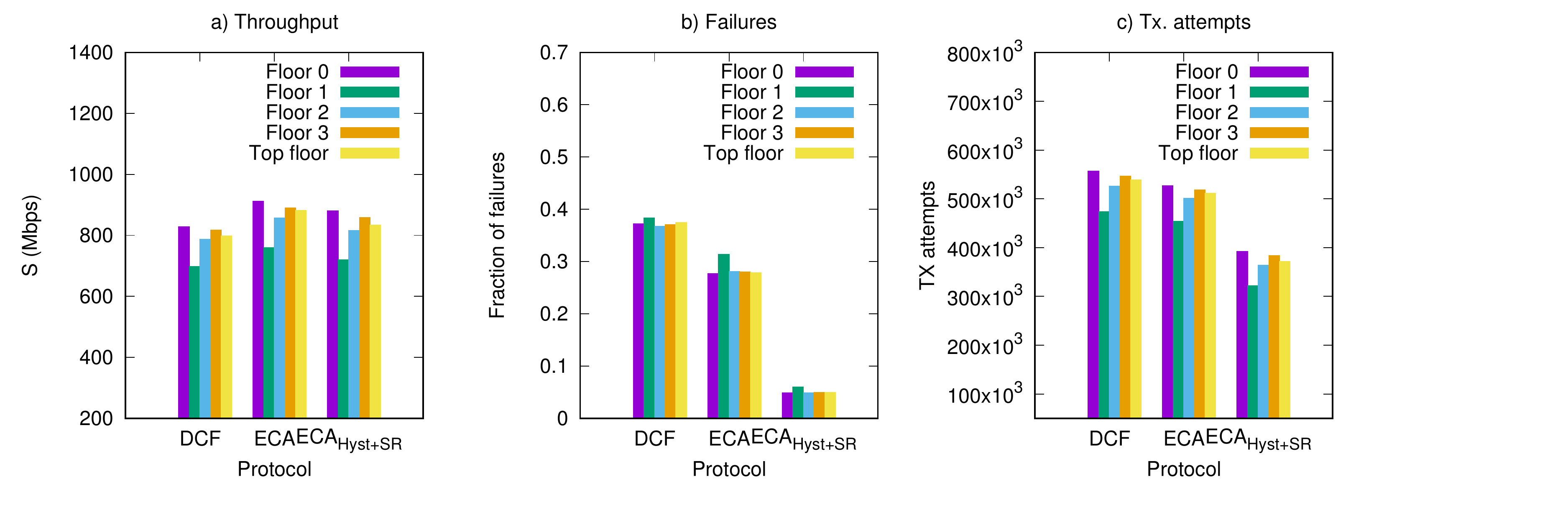}
		\caption{Scenario HEW results using 20 non-overalapping channels which allocation to each WLAN is made at random, showing a) throughput (S), b) fraction of failed transmissions, and c) number of transmission attempts. Each figure shows the aggregat metric per floow, please refer to Figure~\ref{building} for orientation. The walls and floors impose propagation losses presented in Table~\ref{tab1}}
		\label{figCthroughput-segmented-big-random}
	\end{figure}

\item Figure~\ref{figCthroughput-segmented-big-random} shows simulation results for $C=20$, but the allocation is made at random.

This scenario, al-though not ideal, can be considered closer to reality. Figure~\ref{figCthroughput-segmented-big-random} shows higher throughput for CSMA/ECA, but the difference with CSMA/ECA$_\text{Hyst+SR}$ is very small. As the latter still shows lower failures and transmission attempts, it is considered as the overall best in this scenario.

\end{enumerate}

%% file: conclusions.tex
\section{Conclusions}\label{conclusions}

As scenarios become more crowded, the interaction among Overlapping BSS (OBSS) becomes of significant importance for optimising the MAC throughput. The Clear Channel Assessment (CCA) mechanism for determining the state of the channel (busy or empty) relies on MAC-specific\footnote{That is, IEEE 802.11-specific} thresholds, which can change considerably the size of the contention domain of a node.

DCF, as it is based on a random backoff technique, is gravely affected by dense scenarios. Degrading the overall throughput as the number of contenders increases. CSMA/ECA on the other hand, uses a deterministic backoff after successful transmissions technique, which coupled with extensions like Hysteresis and Schedule Reset allows CSMA/ECA$_\text{Hyst+SR}$ to construct collision-free schedules for more contenders.

We tested these three protocols under different scenarios, ranging from single AP, to a complex residential building following TGax specifications for IEEE 802.11ax. Overall, the deterministic backoff technique increases the number of successful transmissions in all of the tested scenarios, outperforming DCF. Furthermore, Hysteresis and Schedule Reset keep CSMA/ECA$_\text{Hyst+SR}$'s failures way lower than the other tested protocols. As this is achieved using longer collision-free schedules, less transmissions attempts are performed, consequently providing a potential reduction in the overall energy consumption.

It is observed that some attributes are beneficial in certain scenarios. For instance, low contention scenarios can draw benefit from the aggressiveness provided by DCF's random backoff mechanism, which leverages the time wasted recovering from failed transmissions. Whereas high contention conditions, like crowded single-AP or multi-AP scenarios may benefit from a deterministic backoff after successful transmissions, such as the used by CSMA/ECA and CSMA/ECA$_\text{Hyst+SR}$. Results from this work evidence the importance of being able to determine the network conditions, and calls for mechanisms able to adapt the MAC protocol accordingly in order to draw benefits.

Different research directions could be derived from this work:
	\begin{itemize}
		\item Consider multi-rate scenarios, which are affected by propagation loss models.
		\item Dynamic ED and CCA threshold adaptation using Dynamic Sensitivity Control~\cite{DSC-survey}.
		\item Schedule Reset and its relation to the sensitivity thresholds, as SR's decisions are based on what is observed in the channel.
		\item Big data analytics. As scenarios get bigger and complex, better data manipulation techniques should be used to interpret what is really happening in the scenario, and then take decisions about which MAC suites best.
	\end{itemize}
	
As all our implementation is open source and freely available at~\cite{CSMA-ECA-NS3,eca-ns3-tutorial}, we encourage other researchers to learn from our experience and start developing tests using complex mobility and propagation models as the ones used in Scenario HEW.

%% file: multiAP.bbl
\begin{thebibliography}{10}

\bibitem{802Standards}
{IEEE Standard for Information Technology - Telecommunications and Information
  exchange between systems. Local and Metropolitan Area Networks - Specific
  requirements}.
\newblock {\em IEEE Std 802.11TM-2012}, page 1646, 2012.

\bibitem{5453868}
O.~Acholem and B.~Harvey.
\newblock {Throughput Performance in Multihop Networks using Adaptive Carrier
  Sensing Threshold}.
\newblock In {\em Proceedings of the IEEE SoutheastCon 2010 (SoutheastCon)},
  pages 287--291, March 2010.

\bibitem{UPC-DSC}
M.~S. Afaqui, E.~Garcia-Villegas, E.~Lopez-Aguilera, G.~Smith, and D.~Camps.
\newblock {Evaluation of Dynamic Sensitivity Control algorithm for IEEE
  802.11ax}.
\newblock In {\em 2015 IEEE Wireless Communications and Networking Conference
  (WCNC)}, pages 1060--1065, March 2015.

\bibitem{barcelo2008lba}
J.~Barcelo, B.~Bellalta, C.~Cano, and M.~Oliver.
\newblock {Learning-BEB: Avoiding Collisions in WLAN}.
\newblock In {\em Eunice}, 2008.

\bibitem{barcelo2011tcf}
J.~Barcelo, B.~Bellalta, C.~Cano, A.~Sfairopoulou, and M.~Oliver.
\newblock {Towards a Collision-Free WLAN: Dynamic Parameter Adjustment in
  CSMA/E2CA}.
\newblock In {\em EURASIP Journal on Wireless Communications and Networking},
  2011.

\bibitem{bellalta2015WCM}
Boris Bellalta.
\newblock {IEEE 802.11 ax: High-Efficiency WLANs}.
\newblock {\em IEEE Wireless Communications (arXiv preprint arXiv:1501.01496)},
  Accepted July 2015.

\bibitem{tuningCarrierSense}
Jing Deng, Ben Liang, and Pramod~K Varshney.
\newblock {Tuning the Carrier Sensing Range of IEEE 802.11 MAC}.
\newblock In {\em IEEE Global Telecommunications Conference, GLOBECOM'04},
  volume~5, pages 2987--2991, 2004.

\bibitem{HEW-scenarios}
{IEEE 802.11 TGax}.
\newblock {TGax Simulation Scenarios}.
\newblock
  {https://mentor.ieee.org/802.11/dcn/14/11-14-0980-16-00ax-simulation-scenarios.docx},
  2014.

\bibitem{HEW}
{IEEE 802.11 TGax}.
\newblock {Status of Project IEEE 802.11ax High Efficiency WLAN (HEW)}.
\newblock {http://www.ieee802.org/11/Reports/tgax$\_$update.htm}, 2016.

\bibitem{JFI}
R.~Jain, D.M. Chiu, and W.R. Hawe.
\newblock {\em A Quantitative Measure of Fairness and Discrimination for
  Resource Allocation in Shared Computer System}.
\newblock Eastern Research Laboratory, Digital Equipment Corporation, 1984.

\bibitem{jamil2014improving}
Irfan Jamil, Laurent Cariou, and Jean-Francois Helard.
\newblock {Improving the capacity of future IEEE 802.11 high efficiency WLANs}.
\newblock In {\em Telecommunications (ICT), 2014 21st International Conference
  on}, pages 303--307. IEEE, 2014.

\bibitem{BECA-test}
{L. Sanabria-Russo}.
\newblock {Report: Prototyping Collision-Free MAC Protocols in Real Hardware}.
\newblock Webpage, 2013.

\bibitem{ns3BuildingDesing}
{NS-3 Project}.
\newblock {Buildings Module}.
\newblock {https://www.nsnam.org/docs/release/3.14/models/html/buildings.html},
  2011.

\bibitem{perahia2013next}
Eldad Perahia and Robert Stacey.
\newblock {\em {"Next Generation Wireless LANs: 802.11 n and 802.11 ac"}}.
\newblock {Cambridge University Press}, 2013.

\bibitem{ns3}
George~F Riley and Thomas~R Henderson.
\newblock {The NS-3 network simulator}.
\newblock In {\em Modeling and Tools for Network Simulation}, pages 15--34.
  Springer, 2010.

\bibitem{ECAqosPaper}
L.~Sanabria-Russo and B.~Bellalta.
\newblock {Traffic Differentiation in Dense Collision-Free WLANs using
  CSMA/ECA}.
\newblock {\em arXiv preprint arXiv:1512.02062}, 2015.

\bibitem{sanabria2014high}
Luis Sanabria-Russo, Jaume Barcelo, Boris Bellalta, and Francesco Gringoli.
\newblock {A High Efficiency MAC Protocol for WLANs: Providing Fairness in
  Dense Scenarios}.
\newblock {\em arXiv preprint arXiv:1412.1395v2}, 2015.

\bibitem{eca-ns3-tutorial}
{Sanabria-Russo, L.}
\newblock {CSMA/ECA in NS-3: a super short tutorial}.
\newblock Webpage, 2016.

\bibitem{CSMA-ECA-NS3}
{Sanabria-Russo, L}.
\newblock {Github repository: CSMA-ECA-NS3}.
\newblock Webpage, accessed June 2016, 2016.

\bibitem{CF-MAC}
{Sanabria-Russo, Luis and Gringoli, Francesco and Barcelo, Jaume and Bellalta,
  Boris}.
\newblock {Implementation and Experimental Evaluation of a Collision-Free MAC
  Protocol for WLANs}.
\newblock Arxiv pre-print, 2014.

\bibitem{DSCNS3}
{Shahwaiz Afaqui, M.}
\newblock {DSC calibration results with NS-3}.
\newblock
  {https://mentor.ieee.org/802.11/dcn/15/11-15-1316-03-00ax-dsc-calibration-results-with-ns-3.pptx},
  2015.

\bibitem{DSC}
{Smith, Graham}.
\newblock {Proposed text for 11ax Draft with respect to Dynamic Sensitivity
  Control (DCS)}.
\newblock
  {https://mentor.ieee.org/802.11/dcn/16/11-16-0310-01-00ax-dsc-proposed-text.docx},
  2016.

\bibitem{stoffers2012comparing}
Mirko Stoffers and George Riley.
\newblock {Comparing the NS-3 Propagation Models}.
\newblock In {\em 2012 IEEE 20th International Symposium on Modeling, Analysis
  and Simulation of Computer and Telecommunication Systems}, pages 61--67.
  IEEE, 2012.

\bibitem{DSC-survey}
C.~Thorpe and L.~Murphy.
\newblock {A Survey of Adaptive Carrier Sensing Mechanisms for IEEE 802.11
  Wireless Networks}.
\newblock {\em IEEE Communications Surveys Tutorials}, 16(3):1266--1293, Third
  2014.

\end{thebibliography}
